\documentclass[twocolumn]{webofc}
\usepackage[varg]{txfonts}   % Web of Conferences font
\begin{document}
\title{Investigating an angular correlation between nearby starburst galaxies and UHECRs with the Telescope Array experiment}

\author{\firstname{Armando} \lastname{di~Matteo}\inst{1}\fnsep\thanks{\email{armando.di.matteo@ulb.ac.be}} \and
        \firstname{Toshihiro} \lastname{Fujii}\inst{2}\fnsep\thanks{\email{fujii@icrr.u-tokyo.ac.jp}} \and
        \firstname{Kazumasa} \lastname{Kawata}\inst{2}\fnsep\thanks{\email{kawata@icrr.u-tokyo.ac.jp}}
        on behalf of the Telescope Array Collaboration
}

\institute{Service de Physique Th\'eorique, Universit\'e Libre de Bruxelles, Brussels, Belgium 
\and
           Institute for Cosmic Ray Research, University of Tokyo, Kashiwa, Chiba, Japan
          }

\abstract{%
  The arrival directions of cosmic rays detected by the Pierre Auger Observatory (Auger) with energies above 39~EeV were recently reported to correlate with the positions of 23 nearby starburst galaxies (SBGs): in their best-fit model, 9.7\% of the cosmic-ray flux originates from these objects and undergoes angular diffusion on a $12.9^\circ$~scale. On the other hand, some of the SBGs on their list, including the brightest one (M82), are at northern declinations outside the Auger field of view. Data from detectors in the northern hemisphere would be needed to look for cosmic-ray excesses near these objects. In this work, we tested the Auger best-fit model against data collected by the Telescope Array (TA) in a 9-year period, without trying to re-optimize the model parameters for our dataset in order not to introduce statistical penalties. The resulting test statistic (double log-likelihood ratio) was $-1.00$, corresponding to $1.1\sigma$ significance among isotropically generated random datasets, and to $-1.4\sigma$ significance among ones generated assuming the Auger best-fit model. In other words, our data is still insufficient to conclusively rule out either hypothesis. The ongoing fourfold expansion of TA will collect northern hemisphere data with much more statistics, improving our ability to discriminate between different flux models.
}
\maketitle
\renewcommand{\citet}[1]{Ref.~\citep{#1}}
\section{The Auger analysis}
We will briefly illustrate the main features of the analysis by the Pierre Auger Collaboration \citep{Aab:2018chp}, which we tried to replicate using Telescope Array data.

\subsection{Assumed sources (23 starburst galaxies)}
\begin{figure}[h]
\centering
\includegraphics[width=0.83\columnwidth,clip]{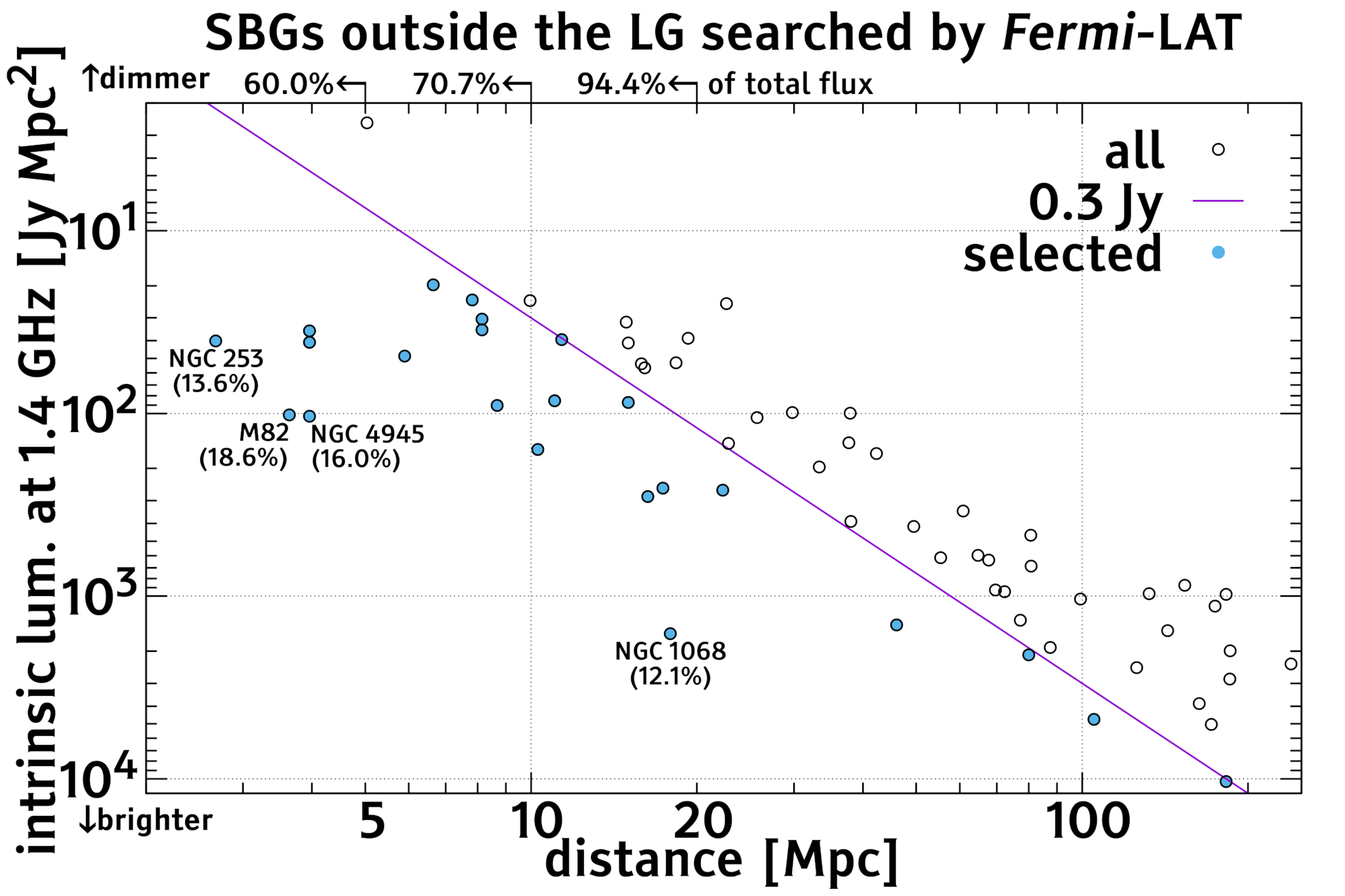}
\caption{Distances from Earth and intrinsic luminosities of objects listed in \citet{Ackermann:2012vca}, with the ones selected in the \citet{Aab:2018chp} analysis highlighted}
\label{fig:galaxies}       % Give a unique label
\end{figure}

In \citet{Aab:2018chp}, the candidate sources were selected from a list of 64 starburst galaxies (SBGs) outside the Local Group
searched by \textit{Fermi}-LAT \citep{Ackermann:2012vca} for gamma-ray emission
(which was only significantly detected from 4 of them in that work).
Among these, \citet{Aab:2018chp} selected the 23 objects whose radio flux at 1.4~GHz is at least 0.3~Jy (figure~\ref{fig:galaxies}).
Their UHECR luminosity was assumed to be proportional to their radio luminosity.

\subsection{The flux model}
\newcommand{\n}{\hat{\mathbf{n}}}
In \citet{Aab:2018chp}, the directional flux of UHECRs from SBGs is assumed to be 
a weighed sum of von Mises--Fisher distributions (the spherical analog of a Gaussian): \[\Phi_\text{SBG}(\n) \propto \sum_\text{sources} \phi_\text{source} \exp(\n \cdot \n_\text{source}/\Psi^2),\] where $\phi_\text{source}$ is taken proportional to the radio flux at 1.4~GHz and $\Psi$ is the RMS deviation in each transverse dimension (the total RMS deviation being $\sqrt{2}\Psi$).  The total flux is assumed to be that from SBGs plus an isotropic background, \[\Phi_\text{model}(\n) = f_\text{SBG}\Phi_\text{SBG}(\n) + (1-f_\text{SBG})\Phi_\text{iso},\] where $\Phi_\text{iso} = 1/4\pi$.

\subsection{The log-likelihood ratio test}
Given two flux models $\Phi_1(\n), \Phi_2(\n)$ and the directional exposure $\omega(\n)$ of the experiment, the log-likelihood ratio test statistic is given by \[\mathrm{TS} = \ln\frac{L_2}{L_1}\text{,~where~}L_i = \prod_\text{events} \frac{\Phi_i(\n_\text{event})\omega(\n_\text{event})}{\int_{4\pi}\Phi_i(\n)\omega(\n)\operatorname{d}\!\Omega}.\] A positive $\mathrm{TS}$ indicates that the data favour the model $\Phi_2$ over $\Phi_1$, and a negative TS vice versa. \citet{Aab:2018chp} used $\Phi_1 = \Phi_\text{iso}$, $\Phi_2 = \Phi_\text{model}$, and searched for the values of the parameters $E_{\min} \in [20~\mathrm{EeV}, 80~\mathrm{EeV}]$, $\Psi$, $f_\text{SBG}$ maximizing $\mathrm{TS}$.

\subsection{Results}
\begin{figure}[h]
\centering
\includegraphics[width=\columnwidth,clip]{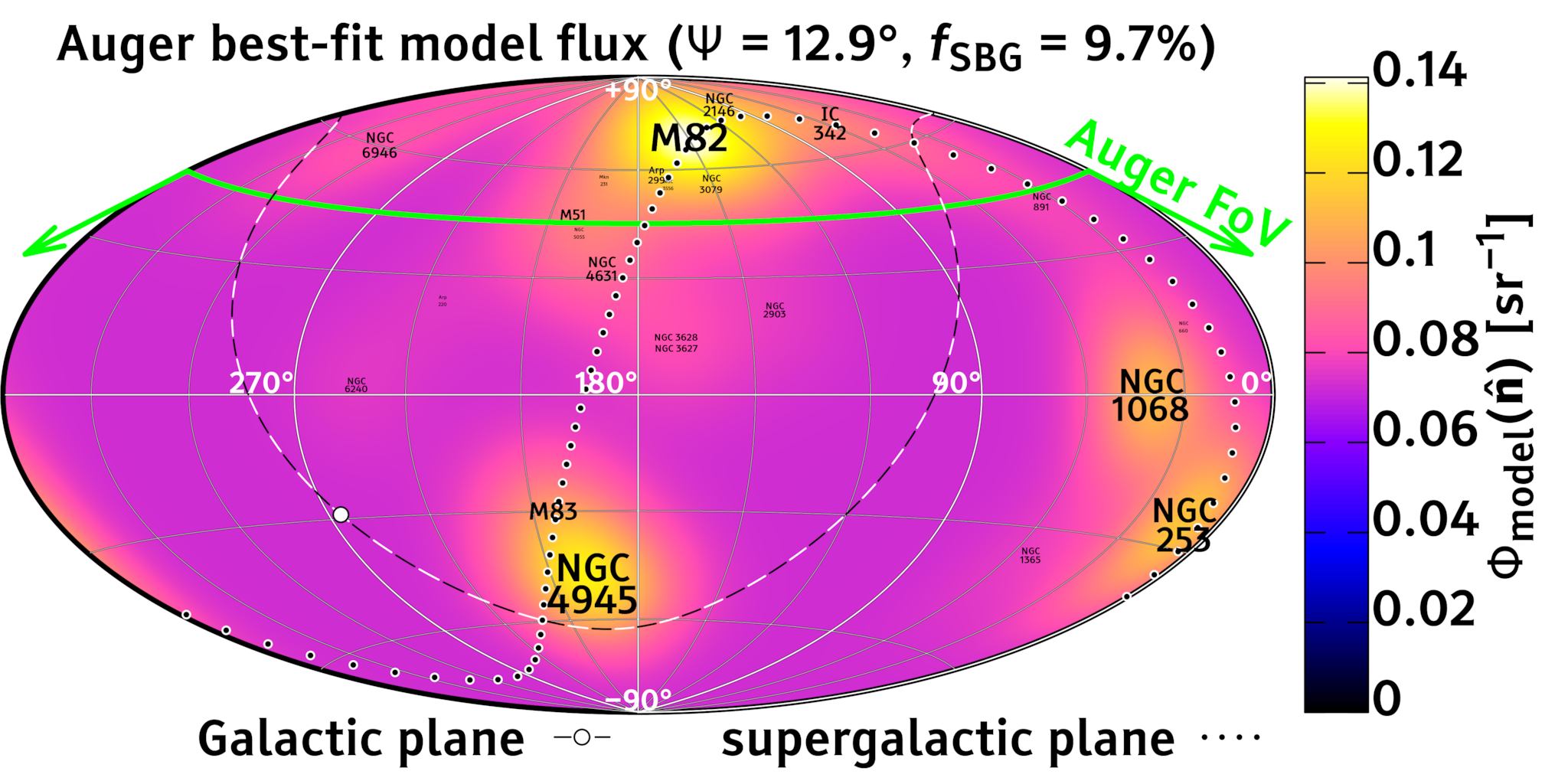}
\caption{The flux map in the best-fit model of \citet{Aab:2018chp}}
\label{fig:skymap}       % Give a unique label
\end{figure}

The best-fit parameters found by \citet{Aab:2018chp} were $\Psi = 12.9^\circ$, $f_\text{SBG} = 9.7\%$, and $E_{\min} = 39$~EeV.  The corresponding flux map is shown in figure~\ref{fig:skymap}.  This model is favoured at the $4.0\sigma$ (post-trial) level over an isotropic null hypothesis and at the $3.0\sigma$ level over the hypothesis that the UHECR emissivity is proportional to the overall matter density in the nearby extragalactic Universe.

\section{The Telescope Array follow-up}
Since several of the objects in the Auger analysis (including the brightest one, M82) are in the Northern Hemisphere outside the Auger field of view,
we decided to test the same hypothesis using Telescope Array data \citep{Abbasi:2018tqo}.

\subsection{The analysis}
We computed TS for our data using the same values of $\Psi$ and $f_\text{SBG}$ as in the best-fit of \citet{Aab:2018chp} ($12.9^\circ, 9.7\%$), without scanning them in order to not introduce any statistical penalty.  We neglected the attenuation of UHECRs in intergalactic space, which was found to be negligible by \citet{Aab:2018chp} because most of the flux from SBGs in this model originates from a few Mpc.

\subsection{The dataset (284 events)}
\begin{figure}[h]
\centering
\includegraphics[width=\columnwidth,clip]{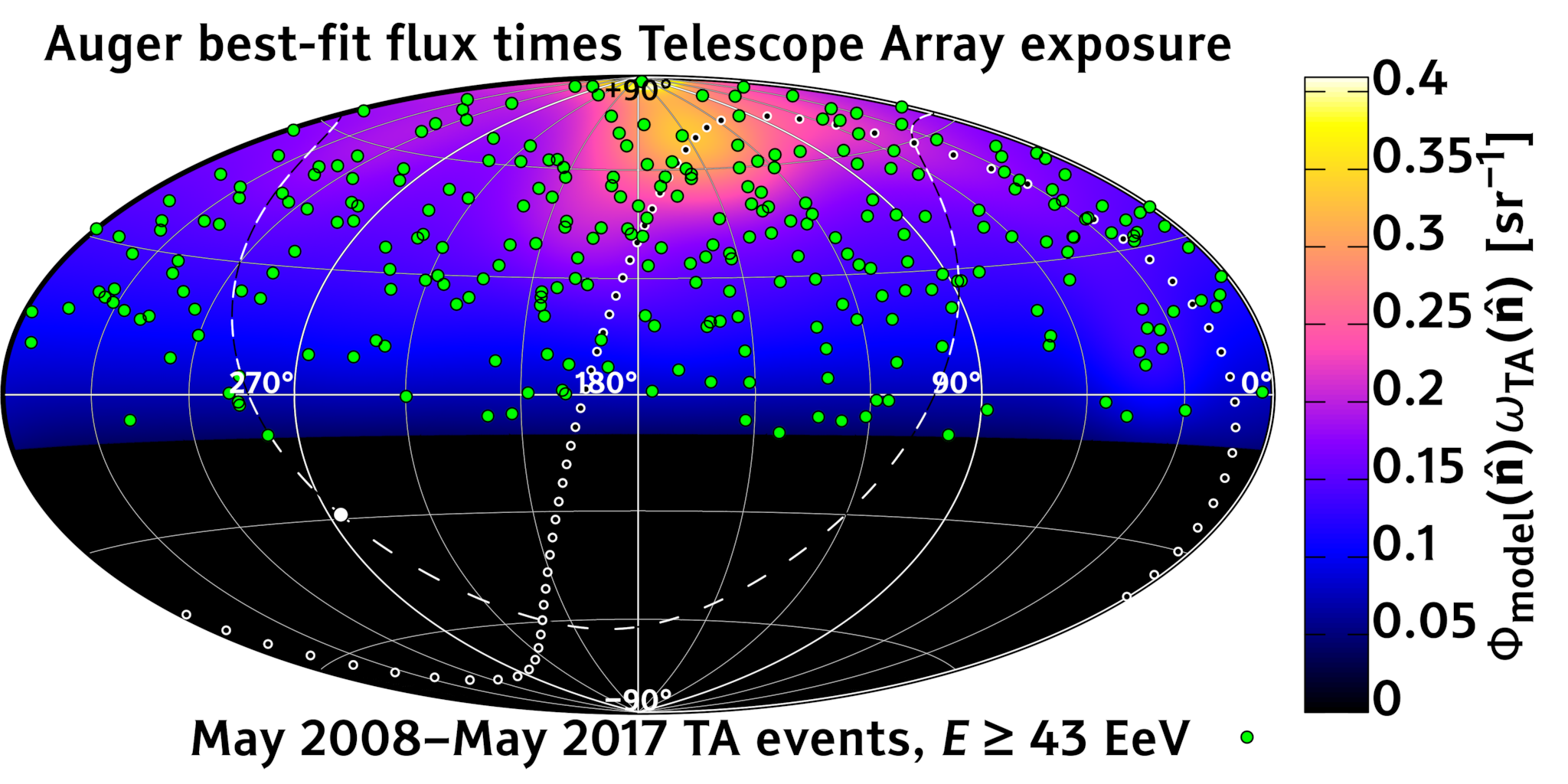}
\caption{The events in the dataset we used and its directional exposure, multiplied by the model flux.}
\label{fig-3}       % Give a unique label
\end{figure}

We used events collected by the Telescope Array ($39.3^\circ$~N, $112.9^\circ$~W, 700~km$^2$~area) in the 9-year period from May 2008 to May 2017. We used the same quality cuts as in \citet{Abbasi:2014lda}, among which zenith angle $\theta \le 55^\circ$ and declination $\delta \ge 10^\circ$. The energy threshold we used is $E_{\min} = 43$~EeV, corresponding to \citet{Aab:2018chp}'s 39~EeV when taking into account the 10\% mismatch in energy scales between the two experiments \citep{Ivanov:2017juh}.  We neglected the measurement resolution ($\lesssim 20\%$ on energy, $\lesssim 1.5^\circ$ on arrival directions) and assumed the detector to be fully efficient in the considered energy and zenith-angle range, calculating its exposure from purely geometrical considerations.  This dataset includes 284 events, whose arrival directions are plotted in figure~\ref{fig-3}, along with the detector exposure multiplied by the model flux.

\subsection{Result and discussion}
\begin{figure}[h]
\centering
\includegraphics[width=\columnwidth,clip]{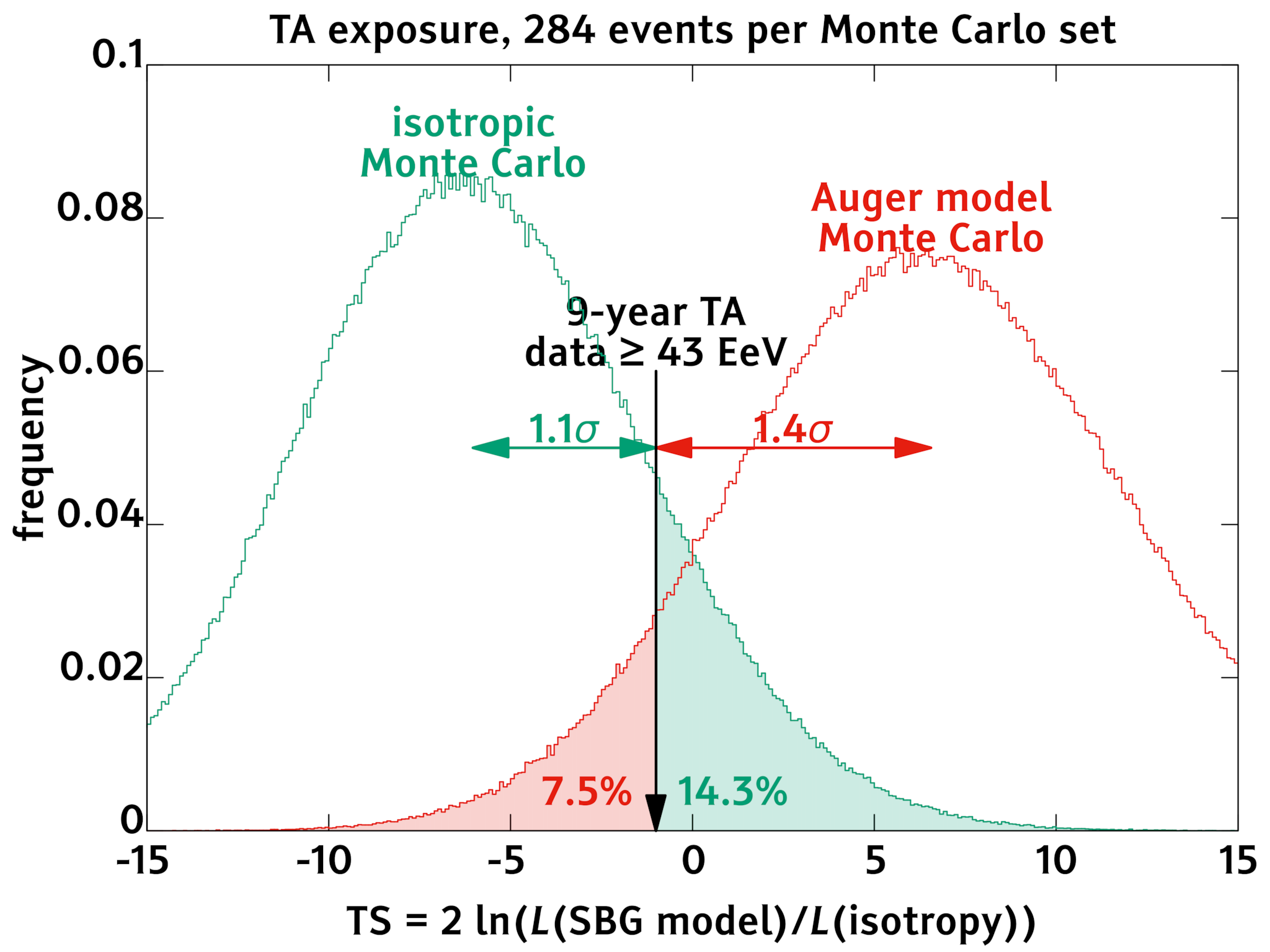}
\caption{The test statistic value we obtained for our data and for Monte Carlo simulations assuming each model}
\label{fig-4}       % Give a unique label
\end{figure}

The test statistic value we obtained was $\mathrm{TS} = -1.00$, less than in 14.3\% of the Monte Carlo simulations we made assuming an isotropic flux but more than in 7.5\% of the simulations assuming the best-fit model from \citet{Aab:2018chp} is correct (figure~\ref{fig-4}). This means our data are still insufficient to rule out either scenario.

The upcoming expansion of Telescope Array, TA$\times$4 \citep{Kido}, will increase its effective area by a factor of 4.  If the data-taking of the TAx4 is started in 2019 as planned, by 2024 it will have collected the equivalent of 30~years of data with the current exposure, improving the statistical sensitivity proportionally to the square root of the number of events\footnote{The separation between the two TS distributions is proportional to the ratio between the standard deviation and the average of each distribution, the former scaling as $\sqrt{N}$ and the latter as $N$.}, hence by a factor~$\approx 1.8$ compared to the present work.

The present study does not include the Local Group objects SMC, LMC, M33 and M31, which were also listed in \citet{Ackermann:2012vca} but in a separate table.  Their intrinsic radio luminosities are very low compared to those in figure~\ref{fig:galaxies} (ranging from 0.19 to 6.2~Jy~Mpc$^2$), but due to their proximities (0.05 to 0.85~Mpc) their radio fluxes are very large (3.3 to 444~Jy), and would dominate the sky if the assumed proportionality between UHECR and radio luminosities also applied to them.  A possible justification for omitting them is hypothesizing that UHECRs are only accelerated during transient events, the expected number of which is proportional to the star-formation rate with a coefficient of order $0.01$--$0.1/(\mathrm{Jy}~\mathrm{Mpc}^2)$, such that Poissonian fluctuations in the brightest few selected objects in figure~\ref{fig:galaxies} are minor but most of the time there are zero events in the Local Group.  One example of such a class of events is gamma-ray bursts from supernovae (one every $\sim 2\,500$~years in M82, each producing UHECRs for $\sim 3\,000$~years) \citep{Biermann:2018clk}.
\bibliography{DIMATTEO_ARMANDO_UHECR2018.bib}

\end{document}